\documentstyle[11pt,newpasp,twoside,epsf]{article}
\def\msun{M$_\odot$}

\def\kms{km\,s$^{-1}$}
\def\halpha{H$\alpha$}

\def\edcomment#1{\iffalse\marginpar{\raggedright\sl#1\/}\else\relax\fi}
\reversemarginpar

\begin{document}
\title{Molecular gas in the intergalactic medium of Stephan's Quintet}

\vspace*{-0.1cm}

 \author{Ute Lisenfeld, St\'ephane Leon}
\affil{Instituto de Astrof\'\i sica de Andaluc\'\i a (CSIC), Apdo. 3004, 
18080, Granada, Spain}
\author{Jonathan Braine}
\affil{Observatoire de Bordeaux, UMR 5804, CNRS/INSU, B.P. 89, F-33270 Floirac,
France}
\author{Pierre-Alain Duc}
\affil{CNRS URA 2052 \& CEA/DSM/DAPNIA Service d'Astrophysique, Saclay, 
91191 Gif sur Yvette Cedex, France}
\author{Vassilis Charmandaris}
\affil{Cornell University, Astronomy Department, Ithaca, NY 14853, USA \&
Chercheur Associ\'e, Observ. de Paris, LERMA, Paris, F-75014, France}
\author{Elias Brinks}
\affil{Dep. de Astron.,
Univ. de Guanajuato, 
Apdo. 144, Guanajuato, Gto 36000, Mexico, \& INAOE, Apdo. 51 \&
216, Puebla, Pue 72000, Mexico
}

\vspace*{-0.2cm}

\begin{abstract}
Stephan's Quintet (SQ) is  a Hickson Compact Group 
well known for
its complex dynamical and star formation history and its
rich intergalactic
medium (IGM).
In order to study the extent, origin and fate of
the intergalactic molecular gas and its relation
to the formation of stars outside galaxies and
Tidal Dwarf Galaxies (TDGs), we mapped with 
the IRAM 30m antenna carbon monoxide (CO) towards
several regions of the IGM in SQ.
In two star forming regions (SQ~A and B), 
situated in very
different environments, we detected
unusually large
amounts of molecular gas ($3.1 \times 10^9$ \msun \ and $7 \times 10^8$ \msun,
respectively), covering an extended area (between 
15 and 25 kpc). In both regions
the CO clouds have  different
properties and may be of a distinct nature.
The integrated CO line of SQ~A is in particular
much wider than in SQ~B. Its CO spectrum 
shows emission  at two velocities (6000 and
 6700 \kms), coincident with
two HI lines, with  
the stronger emission at 6000 \kms \ being very smoothly distributed
without a distinct peak in the starburst region.
In SQ~B the CO emission coincides with 
that of tracers of star formation (\halpha, near-infrared
15 $\mu$m and radio continuum).
The CO peak lies close to the HI peak towards
a steep HI gradient. This is indicating that the
molecular gas is forming in-situ, 
with subsequent star formation taking place. The star forming region
at SQ~B is the object in SQ that most resembles 
a TDG.

\end{abstract}

\section{Galaxy interaction and intergalactic star formation in SQ}

\begin{figure}
\plotfiddle{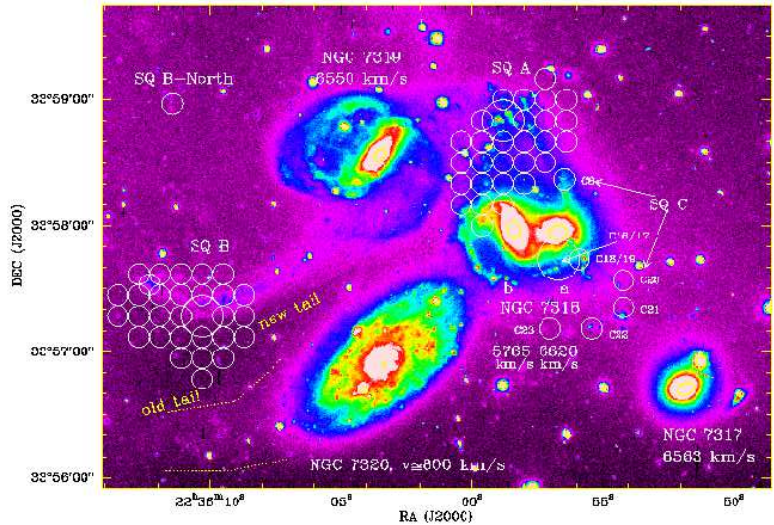}{6.cm}{270}{45}{45}
{-170}{230}
\caption{A V-band image of Stephan's 
Quintet obtained from the CFHT archive. 
NGC~7320 is a foreground galaxy. The fourth
member of the group, NGC~7320c, lies about 4$^\prime$ east of NGC~7319.
The positions observed in CO
are indicated by circles. The large  circle
shows the central (i.e. offset 0,0) position in each region
and  gives the size of the CO(1--0) beam. 
 }
\end{figure}

\begin{figure}
\plotfiddle{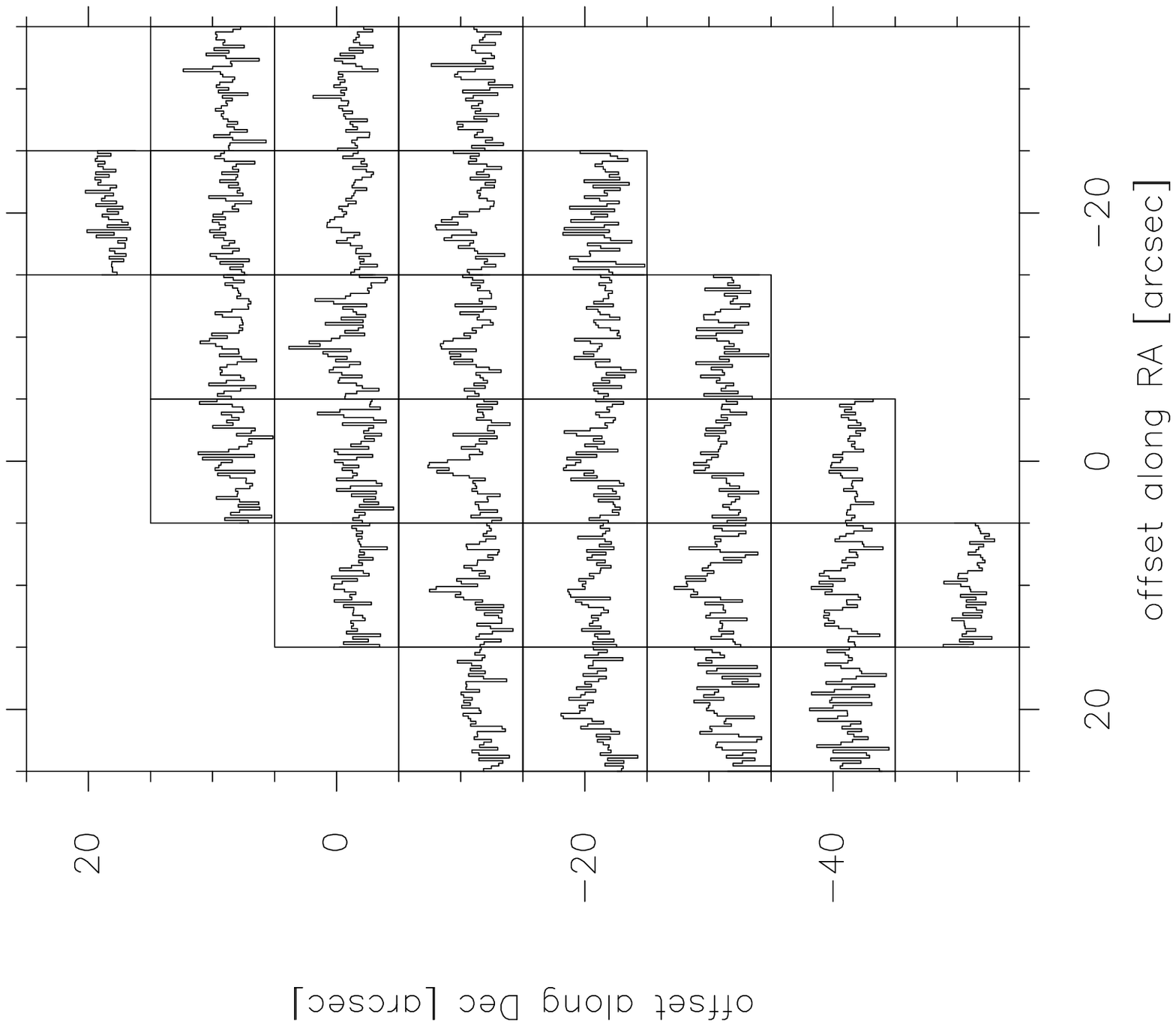}{5.5cm}{270}{45}{45}
{-205}{200}
\plotfiddle{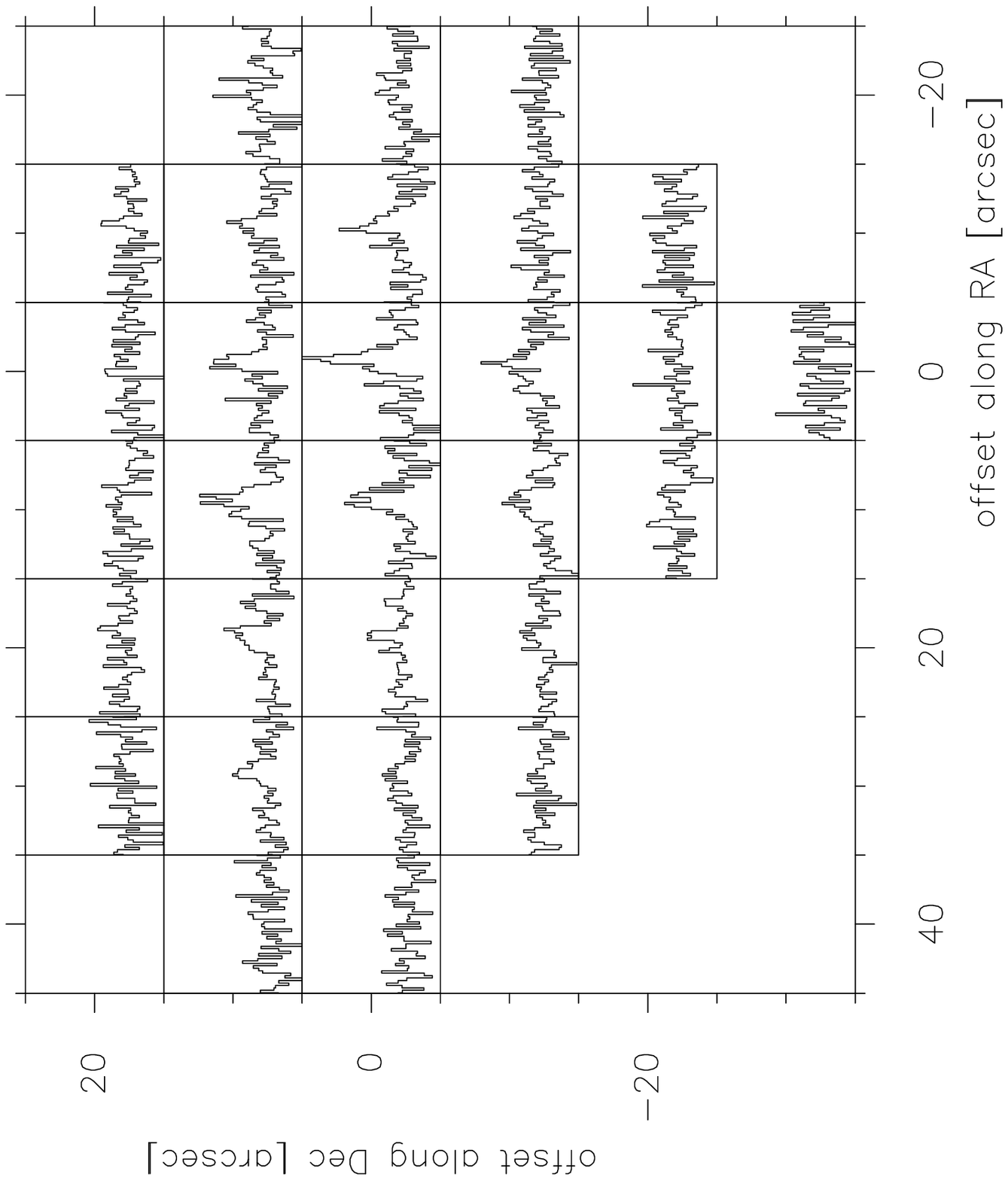}{1.cm}{270}{43}{43}
{-15}{235}
\caption{{\bf Left:} CO(1--0) emission in SQ~A. 
The x-scale in the individual spectra 
is velocity and ranges from 5800 to 6300\,\kms.
The y-scale is in $T_{\rm mb}$ and ranges from -13 mK to 25 mK.
{\bf Right:}
CO(1--0) emission in SQ~B.
The x-scale in the individual spectra 
is velocity and ranges from 6450 to 6750\,\kms.
The y-scale is in $T_{\rm mb}$ and ranges from -13\,mK to 38\,mK.
}
\end{figure}

Stephan's Quintet (Hickson Compact Group 92; hereafter SQ, see Fig. 1)
is one of the best studied examples of a 
Hickson Compact Group. It 
contains four interacting galaxies, NGC~7319,
NGC~7318a, NGC~7318b, and NGC~7317.   
One of its most striking properties is that 
the major part of the atomic gas is in the intragroup medium
(Williams et al. 2002), most
likely the result of interactions in the past and present.

A plausible scenario for the dynamical history of SQ is presented
by Moles, Sulentic \& M\'arquez (1997).
They suggest that
a few times 10$^8$ yr ago the group experienced a collision with
NGC~7320c, a galaxy $\sim$4$^\prime$ to the east of NGC~7319
but with a very similar recession velocity 
(6583 \kms, Sulentic et al. 2001) to the other galaxies 
in SQ.
This collision removed most of the gas of NGC~7319 towards the west 
and east,  and produced the eastern
tidal tail which connects to NGC~7319.  
Presently, the group is
experiencing another collision, this time with 
the ``intruder'' galaxy NGC~7318b,
with a recession velocity of 5765\,\kms,
which is affecting strongly the intergalactic medium (IGM) which
was removed during
the first collision. 
Sulentic et al. (2001), in a multi-wavelength study of the
group, confirm this scenario and suggest that the group has been
visited twice by NGC~7320c. The first collision created the very faint tidal
arm east of the interloper NGC~7320, whereas the second interaction
produced the tidal arm which stretches from NGC~7319 eastwards.

This violent dynamical history has induced star formation at various
places outside the individual galaxies.  ISOCAM mid-infrared and
H$\alpha$ observations have revealed a starburst region (object
A in Xu, Sulentic \& Tuffs 1999, hereafter called SQ~A) 
at the intersection of two faint optical
arms stemming north from NGC~7318a/b. 
Molecular gas has been found at the
position of SQ~A by Gao \& Xu (2000) with BIMA and by Smith \& Struck (2001)
with the NRAO 12m radio telescope.
Both the molecular and the atomic
gas in this location present two velocity components, 
centered at about 6000 and 
6700\,\kms, which implies that they originate from different
galaxies.

A second region where several knots of star
formation are found to lie outside the main
galaxies is  in the tidal arm extending from NGC~7319
to the east. At this position, identified as B by (Xu et al. 1999) and
hereafter called SQ~B, there is also mid-infrared and H$\alpha$
emission, although much weaker than in SQ~A. 
Braine et al. (2001) in a study of the CO emission of 
TDGs detected 2.9$\times$10$^8$\,\msun \  of molecular gas
at the position of SQ~B. 
In order to follow up on these detections and to study the extent, 
origin and fate of the CO in the intergalactic medium (IGM) in SQ, we  
carried out the single dish survey presented here.

\section{Intergalactic molecular gas in SQ}

\begin{figure}
\plottwo{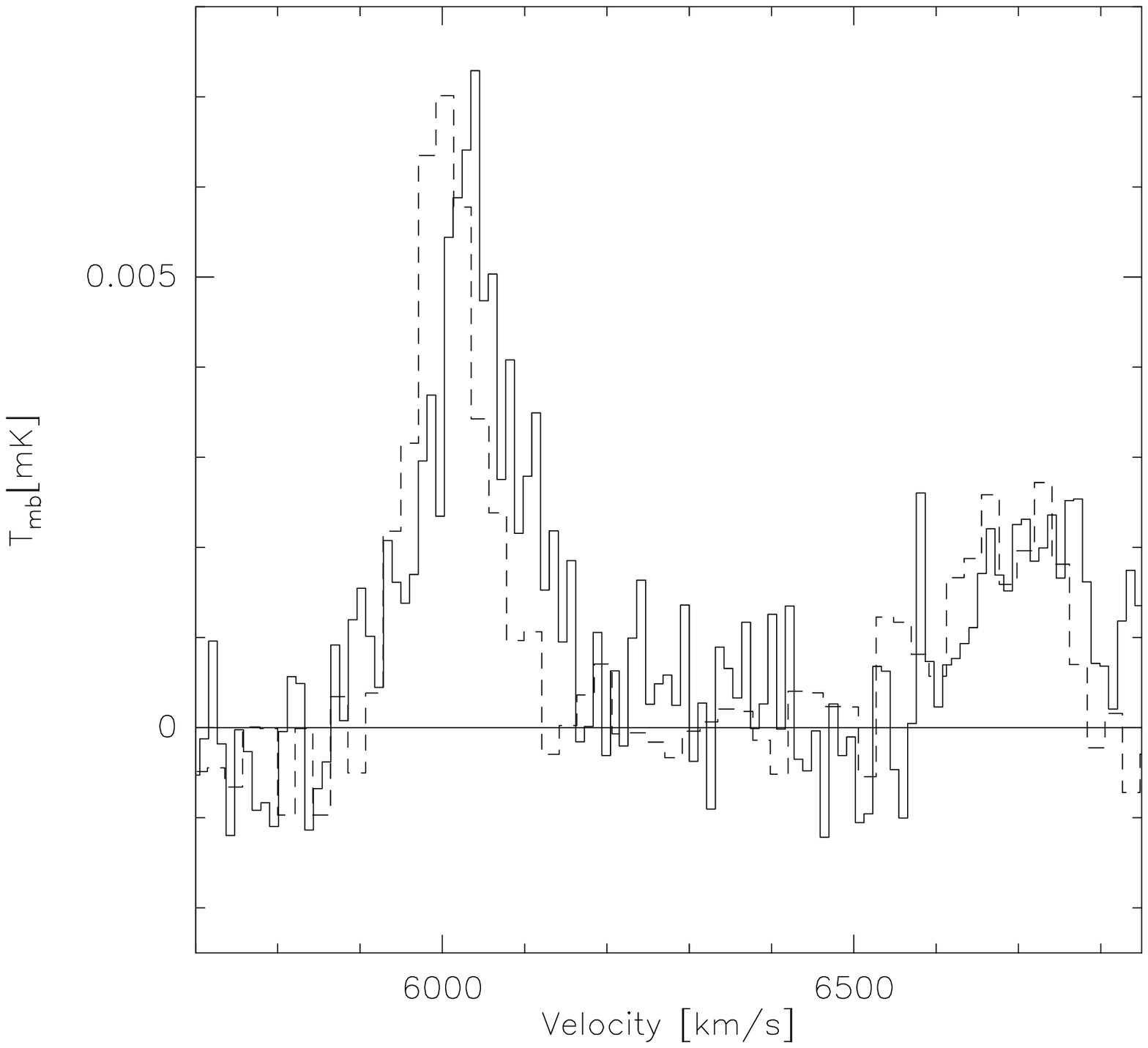}
{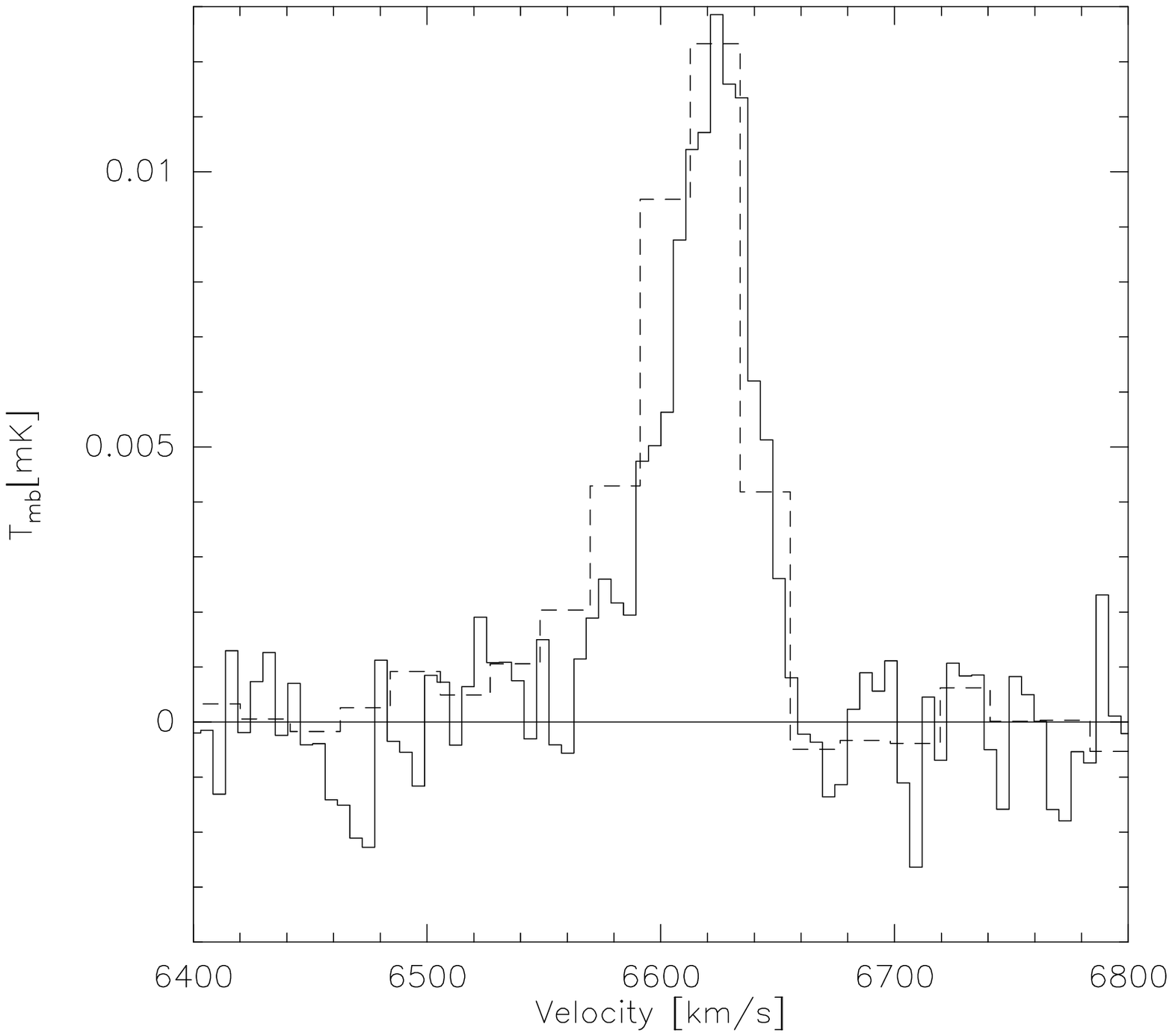}
\caption{{\bf Left:} 
CO(1--0) (full line) and HI (dashed line, in arbitrary units,
from Williams et al. 2002) spectrum of SQ~A, 
averaged over  the total observed area. 
{\bf Right:}
CO(1--0) (full line)  and HI (dashed line, in arbitrary units,
from Williams et al. 2002) spectrum of SQ~B,    
averaged over the central 15 positions.
}
\end{figure}

\begin{figure}
\plottwo{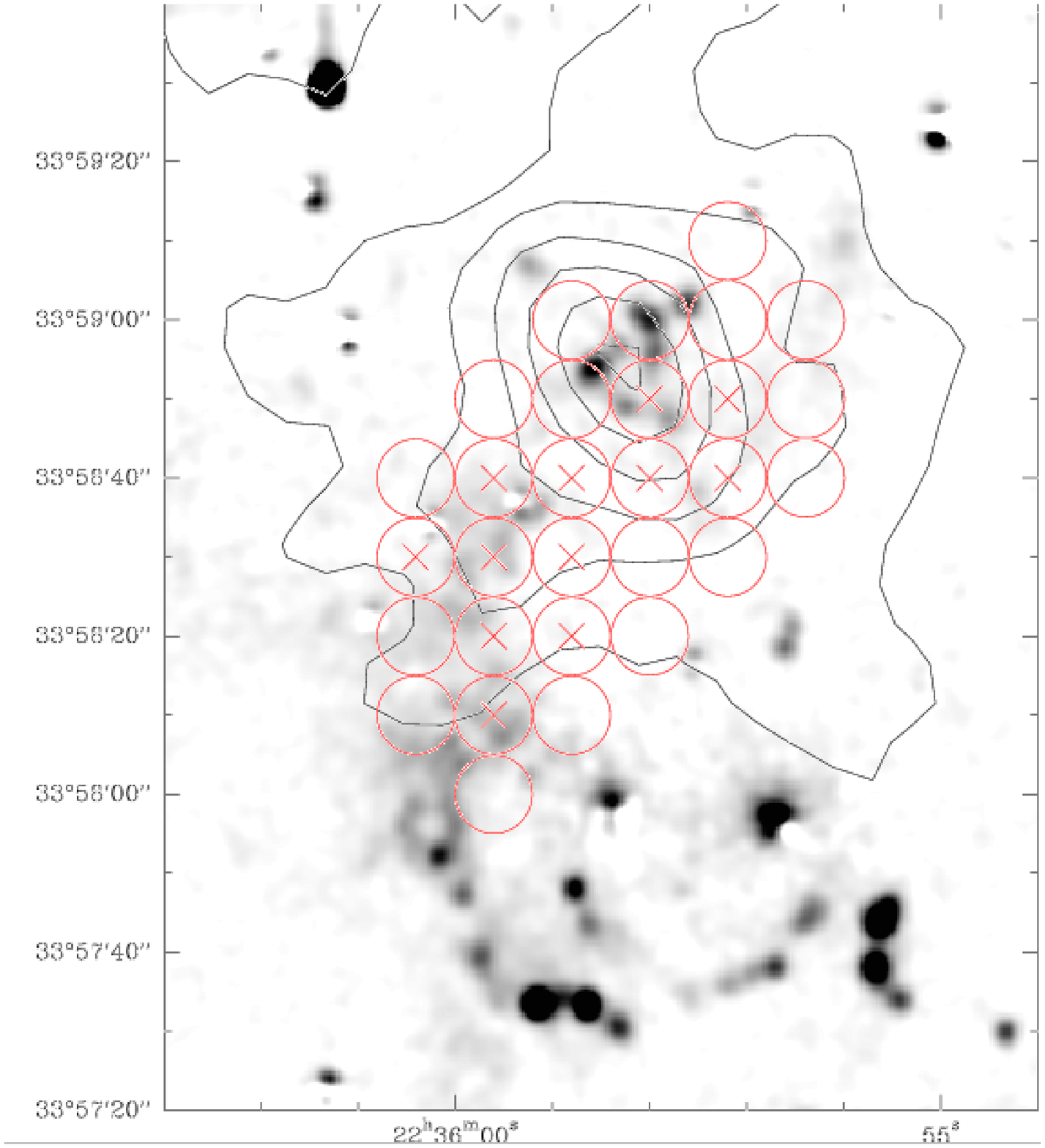}
{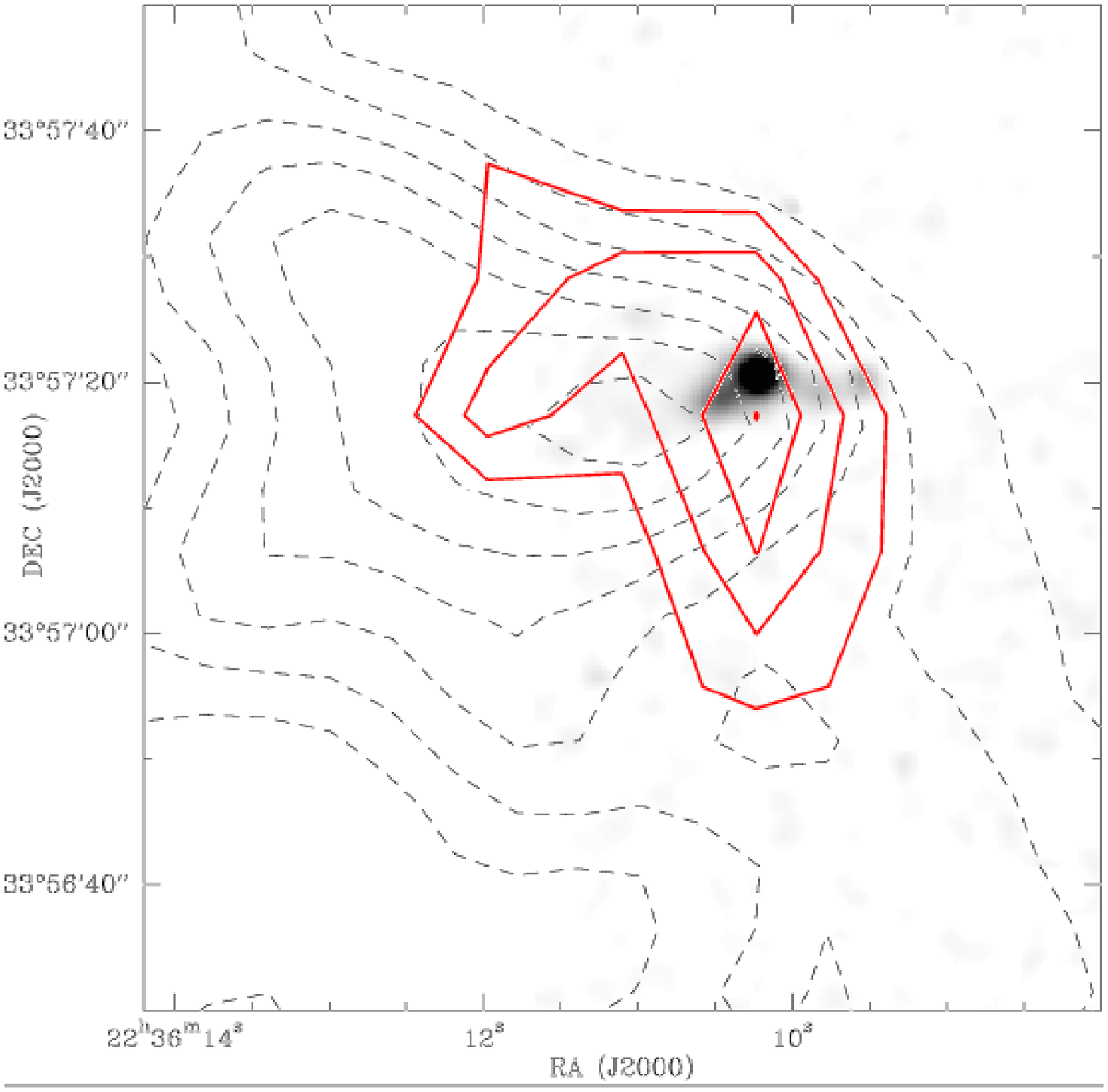}
\caption{{\bf Left:} A grey scale image of the \halpha \, emission 
towards SQ~A centered at a velocity of 6000 \kms \
(from Xu et al. 1999), 
overlaid with a velocity integrated HI contour map covering the
same velocity (Williams et al. 2002).
The circles show the locations of our pointings and the crosses indicate the
positions where CO emission was detected above a level of $ 3 \sigma$.
{\bf Right:}
A grey scale image of the \halpha \ emission towards SQ~B
centered at a velocity of 6600 \kms \
(from Xu et al. 1999), 
overlaid with a velocity integrated HI contour map covering the
same velocity (Williams et al. 2002,
dashed contours),  and
the velocity integrated CO emission, $I_{\rm CO}$, 
(full contours,
levels at 0.4, 0.65, 0.9 and  1.15 K km s$^{-1}$).
 }
\end{figure}

We mapped SQ in July 2001 with the IRAM 30m telescope in CO(1--0) and
CO(2--1). 
Details of the observations and a more 
complete discussion of the results can be found in Lisenfeld et al. (2002).
The positions observed are shown in Fig. 1.
We mapped (i) the region around the intergalactic starburst SQ~A, 
(ii) the area around the star forming region along the 
eastern tidal tail (SQ~B) and (iii)  several positions around
NGC~7318a/b observed by Mendes de Oliveira et al. (2001) in  \halpha \
(labelled as SQ~C in Fig. 1) and suggested to be candidates for TDGs.
At these latter  positions no molecular gas was found.

Both in SQ~A and SQ~B we found extended and abundant molecular gas (see
Fig. 2), the molecular gas masses being 
$3.1 \times 10^9$ \msun \ and $7 \times 10^8$ \msun,
respectively, and with spatial extents between 15 and 25 kpc. 
The molecular-to-atomic mass ratio is 1.2 (0.5) for SQ~A (SQ~B), much
higher than typical values found in dwarf and tidal dwarf galaxies
(Braine et al. 2001). The 
integrated intensity ratios are similar (CO(2-1)/CO(1-0)=$0.69 \pm 0.16$,
respectively $0.56\pm 0.13$), consistent with optically thick emission.
In both regions, the CO and HI
velocities and line shapes agree very well (Fig. 3), although in SQ~A the
CO velocity seems to be shifted slightly to higher velocities with
respect to the HI. In SQ~A, the CO spectrum shows two velocity components,
one at about 6000 \kms (the velocity of the intruder galaxy NGC~7318b) and
the other at 6700 \kms \ (the velocity of the rest of the SQ galaxies). Only
the component at 6000 \kms \ is visible in the individual spectra.


\section{SQ~A and SQ~B -- two completely different star-forming regions}

SQ~A and SQ~B are situated in completely different environments:
SQ~A is part of the region presently affected by the collision with
NGC~7318b and hosts shocks as evidenced  through X--ray, radio continuum
and \halpha \ emission.
SQ~B is part of the tidal tail situated in a more quiescent area. 
Therefore it is not surprising that in spite of their
similarities (the large amount and extent of the
molecular gas, the high molecular-to-atomic
gas mass ratio and the similar CO(2--1)/CO(1--0) line ratio) other
properties of the molecular gas in SQ~A and SQ~B are fundamentally
different.

The lines in SQ~A at  6000 \kms \ are very
broad (FWHM 60--80 \kms, see Fig. 3), reflecting the  wide range of velocities
present in the IGM affected by the collision. The CO distribution 
is rather smooth, spatially offset from the HI emission,  
with no distinct peak and with only weak emission at the location
of the starburst 
(in Fig. 4 the starburst is situated at the peak of the HI emission
coincident with  \halpha \ knots).
Deriving the dynamical mass via a simple estimate from the 
linewidth and spatial extent, we derive 
$1.9 \times 10^{10}$ \msun, much larger than the
gas and stellar mass in this area ($\sim 3.8 \times 10^9$ \msun), 
suggesting that  the molecular gas cloud is not
gravitationally bound on the size scales at the current resolution.
Thus, in SQ A, the extended and homogeneous distribution of the molecular gas
makes it unlikely that a gravitational collapse is responsible for 
its formation.
The situation might be different for the emission at 6700 \kms, for which
we cannot constrain the spatial extent 
because of the weakness of the lines.  
However, assuming that it closely follows 
the HI emission at this velocity, 
which is very concentrated around the starburst, it is likely to be more 
restricted to the starburst region.

In SQ~B, on the other hand, the lines are much
narrower (FWHM 30--40 \kms) and the CO distribution  
peaks at the star forming region in the tidal arm, close to the
peak of the HI emission (Fig. 4).
The dynamical mass that can be derived from the CO emission 
$(2.9 \times 10^{9}$ \msun) is
comparable to the total gas mass  ($2.1 \times 10^{9}$ \msun)
and suggests that the molecular gas
is gravitationally bound. 
All this indicates that we are seeing 
the formation of molecular gas from atomic gas and 
subsequent star formation.
This finding, together
with the position of SQ B on the tidal tail and the similarity
of its properties to other dwarf galaxies, make it the best
candidate for a TDG in SQ.

\section{Other cases of intergalactic molecular gas}

The first intergalactic molecular clouds which were discovered by
Brouillet, Henkel  \& Baudry (1992) as part of the M~81 group tidal material.
Later, a similar object was detected in the same group by
Walter \& Heithausen (1999). In both cases, no optical counterpart
was found. These clouds, with masses of less than $5\times 10^7$ \msun,
might be the first stages of TDGs.

Smith et al. (1999) detected $4\times 10^8$ \msun \ in the tidal
tail of Arp 215.
In a survey of molecular gas in TDGs 
Braine et al. (2000) and Braine et al. (2001)
detected 8 of 11 objects and derived
molecular gas clouds between several $10^6$ and
$3\times 10^8$ \msun. 

These studies show that intergalactic molecular gas in tidal features is 
a normal phenomenon, in which molecular gas is sometimes detected 
at distances of up to
$\sim 100$ kpc from the parent galaxies. However, the enormous quantities
and large extent of molecular gas found in SQ are unprecedented.


\begin{references}
\reference Braine, J.,  Lisenfeld, U., Duc, P.-A., 
	Leon, S. \& Brinks, E. 2000, Nature, 403, 867
\reference Braine, J., Duc, P.-A., Lisenfeld, U., Charmandaris, V.,
	Vallejo, O., Leon, S. \&  Brinks, E. 2001, \aap, 378, 51
\reference Brouillet, N., Henkel, C. \&  Baudry, A. 1992, \aap, 262, L5
\reference Gao, Y. \&   Xu, C. 2000, \apj, 542, L83
\reference Lisenfeld, U.,  Braine, J., Duc, P.-A., Leon, S., Charmandaris, V.
\& Brinks, E. 2002, \aap, in press, astro-ph/0208494
\reference Mendes de Oliveira, C., Plana, H., Amram, P., Balkowski, C. \&  
Bolte, M. 2001, \aj, 121, 2524
\reference Moles, M., Sulentic. J.W. \&  M\'arquez, I. 1997, \apj, 485, L69
\reference Smith, B.J., Struck, C., Kenney \&  J.D.P, Jogee, S. 1999, 
\aj, 117, 1237
\reference Smith, B.J. \&  Struck, C. 2001, \aj, 121, 710
\reference Sulentic, J. W., Rosado, M., Dultzin-Hacyan, D.,
	Verdes-Montenegro, L., Trinchieri, G., Xu, C. \&  Pietsch, W.
	2001, \aj, 122, 2993
\reference Walter, F. \&  Heithausen, A. 1999, \apj, 519, L69
\reference Williams, B.A., Yun, M.S., Verdes-Montenegro, L. \&  van Gorkom,
	J.H. 2002, \aj, 123, 2417
\reference Xu, C., Sulentic, J.W. \& Tuffs, R. 1999, ApJ, 512, 178
\end{references}
\end{document}